\documentclass[twocolumn]{aastex631}
\usepackage[utf8]{inputenc}
\usepackage[normalem]{ulem}  
\usepackage[T1]{fontenc}
\usepackage{ae,aecompl}
\usepackage{graphicx}
\usepackage{amsmath}
\usepackage{amssymb}
\usepackage{supertabular}
\usepackage{hyperref}
\usepackage{xcolor}
\usepackage{academicons}
\usepackage{ragged2e}
\usepackage{tabularx}
\usepackage{multirow}
\usepackage{booktabs}
\usepackage{threeparttable} 
\usepackage{times}
\usepackage{csvsimple}

\begin{document}
\title{Internal superfluid response and torque evolution in the giant glitch of PSR J1718--3718
}

\author{Peng Liu}
\altaffiliation{These authors contributed equally to this work.}
\affiliation{Department of Physics and Astronomy, Qinghai University, Xining, 810016, China}
\affiliation{Department of Astronomy, Xiamen University, Xiamen 361005, China}
\affiliation{Xinjiang Astronomical Observatory, Chinese Academy of Sciences, 150 Science 1-Street, Urumqi 830011, China}

\author[0000-0001-6836-9339]{Zhonghao Tu}
\altaffiliation{These authors contributed equally to this work.}
\affiliation{Department of Astronomy, Xiamen University, Xiamen 361005, China; liang@xmu.edu.cn}

\author[0000-0002-5381-6498]{Jianping Yuan}
\affiliation{Xinjiang Astronomical Observatory, Chinese Academy of Sciences, 150 Science 1-Street, Urumqi 830011, China}
\affiliation{Xinjiang Key Laboratory of Radio Astrophysics, 150 Science1-Street, Urumqi 830011, China}
\affiliation{School of Astronomy, University of Chinese Academy of Sciences, Beijing 100049, China}

\author[0000-0002-0786-7307]{Ang Li}
\affiliation{Department of Astronomy, Xiamen University, Xiamen 361005, China; liang@xmu.edu.cn}

\correspondingauthor{A. Li}
\email{liang@xmu.edu.cn}

\begin{abstract}
We investigate the post-glitch rotational evolution of pulsars by analyzing the 2007 giant glitch of PSR J1718$-$3718 using a vortex creep model that incorporates both inward and outward nonlinear vortex motion, along with a time-varying external torque. A comprehensive fitting framework is developed, constrained by prior knowledge of moment of inertia participation from previous glitch studies. We apply a Markov Chain Monte Carlo approach to quantify uncertainties and parameter correlations. The model reproduces the observed timing data and yields physically consistent values for moment of inertia fractions and creep timescales.
Our results indicate that inward creep and a long-term change in external torque dominate the observed increase in spin-down rate, pointing to structural changes within the star—likely triggered by a crustquake that initiated both vortex motion and a change in the moment of inertia. We estimate that the glitch involved approximately $2.4 \times 10^{12}$ inward-moving vortices and $\sim142$ crustal plates with a typical size of $\sim0.03$\,km. 
This study demonstrates that detailed post-glitch modeling of sparse timing data can simultaneously constrain internal superfluid dynamics and external torque evolution, providing a quantitative framework to probe the structural properties of neutron star interiors.
\end{abstract}


\section{Introduction} \label{sec:intro}

\begin{figure*}[ht!]
\centering
\includegraphics[width=18 cm]{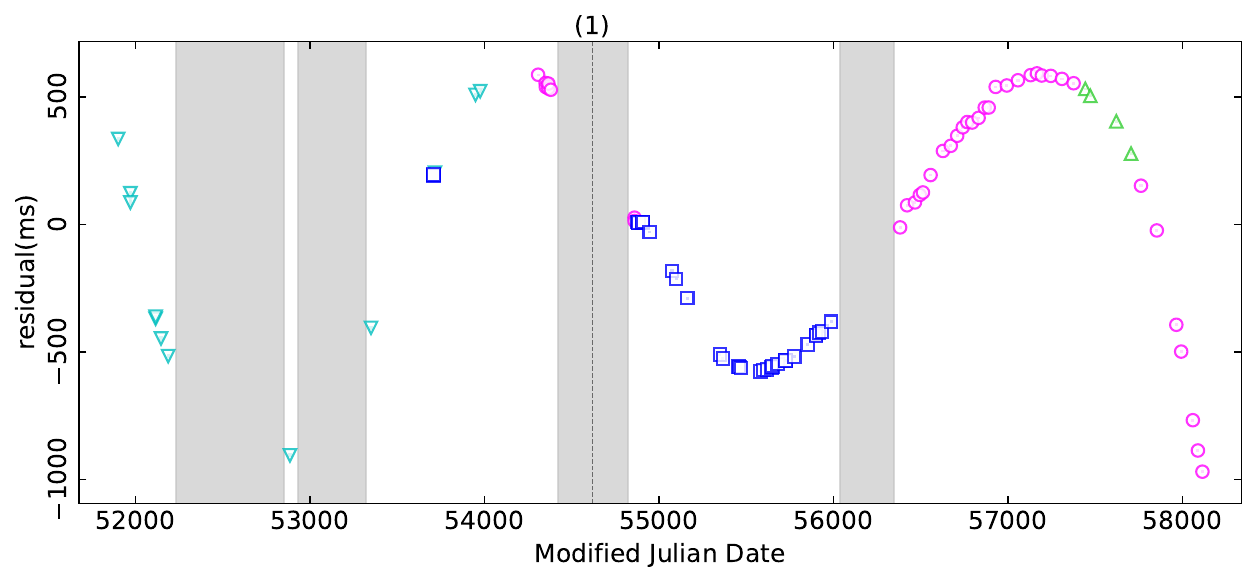}
\caption{
Timing residuals for PSR J1718$-$3718 during MJD 51901--58115, obtained from a timing model incorporating a single glitch. The circles, inverted triangles, triangles, and squares represent the observation data obtained by receivers with center frequencies of 1369 MHz, 1374 MHz, 1465 MHz, and 3100 MHz, respectively.  
The glitch epoch is marked with a vertical dashed line, with the number in parentheses at the top indicating its sequence number. The gray shaded areas indicate periods of data gaps longer than one year.
\label{fig:residual} 
}
\end{figure*}

PSR J1718$-$3718 is a young pulsar with remarkable rotational characteristics. It has a long spin period of $P\sim 3.3786\,\mathrm{s}$ and a large spin-down rate of $\dot{P}\sim 1.6\times 10^{-12}\,\mathrm{s}\,\mathrm{s}^{-1}$ \citep{ManchesterH2011}. These parameters imply a characteristic age of $\tau_{\mathrm{c}}=P/(2\dot{P})\sim 33.2\,\mathrm{kyr}$ and an extraordinarily high surface dipole magnetic field of $B_{\mathrm{s}}=3.2\times 10^{19}\sqrt{P\dot{P}}\sim 7.4\times 10^{13}\,\mathrm{G}$. This places PSR J1718$-$3718 as the rotation-powered pulsar with the second-strongest known magnetic field, approaching the regime of magnetars and even exceeding the field strength of several known magnetars~\citep{OlausenK2014}. 

The pulsar was first discovered in the Parkes multibeam pulsar survey \citep{HobbsFSC2004}. Its magnetar-like nature was further supported by the detection of a coincident, purely thermal X-ray source with the Chandra X-ray Observatory \citep{KaspiM2005}, a suggestion later confirmed when \citet{ZhuKMP2011} showed the X-ray pulse period was consistent with the radio period of PSR J1718$-$3718. 

A defining event in the history of this pulsar was a giant glitch occurring around MJD 54610(244) \citep{ManchesterH2011}. This event had a fractional change in spin frequency of $\Delta\nu/\nu \sim 33.25(1)\times 10^{-6}$, representing the largest amplitude glitch detected in the radio band to date and the third largest overall. What makes this glitch exceptionally rare, however, is its aftermath: unlike the typical exponential decay of the spin-down rate back towards its pre-glitch value, the absolute value of the spin-down rate ($|\dot{\nu}|$) of PSR J1718$-$3718 was observed to \textit{increase} over the following two years \citep{ManchesterH2011}.

In this paper, we present a systematic study of the rotational behavior of PSR J1718$-$3718 based on timing data from Murriyang, the Parkes radio telescope spanning 2000 to 2017. We confirm the unique post-glitch recovery and update the glitch epoch to MJD 54620(240). The central aim of this work is to explain this peculiar behavior within the framework of the vortex creep model \citep{1984ApJ...276..325A,1989ApJ...346..823A}, incorporating both inward and outward vortex motion and a time-varying external torque. Our analysis provides quantitative constraints on the superfluid dynamics and structural changes within the neutron star.

The structure of this paper is arranged as follows: 
Section \ref{sec:Observations} describes the observational data and reduction procedures; Section \ref{sec:result} presents the timing solutions and the anomalous post-glitch recovery; Section \ref{sec:fits} introduces the vortex creep model with inward motion, details the MCMC fitting methodology, and presents the physical interpretation of the results; and Section  \ref{sec:Conclusions} summarizes our findings and their implications.

\begin{table*}
\small
\begin{minipage}[]{90mm}
\caption{Pre- and post-glitch timing solutions for PSR J1718$-$3718. 
\label{tab:int}} 
\end{minipage}
\begin{flushleft}  
\hspace*{-\columnsep}  
  \begin{threeparttable}
    \renewcommand{\arraystretch}{1.1}
    \setlength{\tabcolsep}{10pt}    
    \resizebox{1.05 \textwidth}{!}{
\begin{tabular}{lcc}
  \hline   \hline
Parameter                       & Pre-glitch 1  & Post-glitch 1   \\ 
\hline
Pulsar name (J2000)             & \multicolumn{2}{c}{PSR J1718$-$3718$^{a}$} \\
Right ascension (J2000) (h:m:s) & \multicolumn{2}{c}{17:18:09.83(1)$^{b}$}\\
Declination (J2000) (d:m:s)     & \multicolumn{2}{c}{$-$37:18:51.5(2)$^{b}$}\\ 
DM (cm$^{-3}$\,pc)              & \multicolumn{2}{c}{371.1(17)$^{c}$}\\ 
\hline
Pulse frequency, $\nu$ (Hz)     & 0.29599855892(9)  & 0.29596773419(5) \\
Pulse frequency derivative, $\dot{\nu}$ ($10^{-13}$\rm\ s$^{-2}$) 
                                & $-$1.39938(1)     & $-$1.414183(5) \\
Pulse frequency second derivative, $\ddot{\nu}$ ($10^{-24}$\rm\ s$^{-3}$) 
                                & 0.37(5)            & $-$0.52(2) \\
Epoch of frequency determination (MJD)  &53140    &56486\\
Data span (MJD)  & 51901--54380       & 54859--58115\\
ToA numbers                             &23      &68\\
RMS timing residual ($\rm \mu$s)     &17953     &42639\\
\hline
Time units           &  \multicolumn{2}{c}{TT(TCB)}   \\
Reference time scale &  \multicolumn{2}{c}{TT(TAI)}   \\
Solar System ephemeris model &  \multicolumn{2}{c}{DE440}   \\
  \hline   \hline
\end{tabular} }
\end{threeparttable}
\end{flushleft}%
\begin{tablenotes}
\item[ ] \textit{Note}. References: 
$^a$ \citep{HobbsFSC2004}; 
$^b$ \citep{ZhuKMP2011}; 
$^c$ \citep{ManchesterH2011}.
\end{tablenotes} 
\end{table*}

\section{Observations and Data analysis} 
\label{sec:Observations}
The timing data for PSR J1718$-$3718 used in this paper were obtained from the online domain of Parkes pulsar data archive\footnote{\href{https://data.csiro.au/domain/atnf}{https://data.csiro.au/domain/atnf}}. 
This dataset was collected with Murriyang, the Parkes radio telescope between 2000 and 2017, with most observations carried out using the Parkes multibeam receiver.
The observation system can be roughly divided into two phases: from 2000 to 2006, observations were conducted with a central frequency of 1374 MHz and a bandwidth of 288 MHz, with data recorded using the analogue filter bank (AFB), and integration time was 10--20 mins \citep{ManchesterLCB2001}; 
from 2007 to 2017, the observing setup was adjusted to a central frequency of 1369 MHz with a bandwidth of 256 MHz, and data were collected using the Parkes Digital FilterBank system (PDFB1, PDFB3 and PDFB4) \citep{StaveleyWBD1996,HobbsMDJ2020}, with observing intervals of 2--4 weeks and integration times of 2--15 mins.
In order to enhance the integrity of the dataset, this study supplemented observation data from H-OH receiver and 10 cm receiver.
The H–OH receiver operated at a central frequency of 1465 MHz with a bandwidth of 512 MHz, with data recorded using PDFB4 and integration time was 10--15 mins \citep{DaiJWK2018}.
The 10 cm receiver has a center frequency of 3100 MHz and a bandwidth of 1000 MHz. Its observation data is recorded by PDFB1, PDFB3, and PDFB4, with a typical integration time of 10 mins \citep{GranetZFG2005}.

After obtaining the raw observational data, we processed it offline using the \texttt{PSRCHIVE} software suite \citep{HotanSM2004,StratenDO2012}. The procedure consisted of the following steps:
\begin{itemize}
    \item \textbf{Pre-processing:} Raw data underwent incoherent de-dispersion and manual excision of radio frequency interference (RFI).
    \item \textbf{Profile Generation:} For each observation, an average pulse profile was generated.
    \item \textbf{Standard Profile Creation:} Using \texttt{PSRADD}, average profiles from the same observing frequency were superimposed to create a single, high signal-to-noise ratio standard profile for each frequency band.
    \item \textbf{Time-of-Arrival (ToA) Calculation:} The Fourier-domain Markov chain method \citep{HotanSM2004} was used to cross-correlate each individual average profile with the corresponding standard profile, yielding precise ToAs at the observatory.
    \item \textbf{Final Dataset:} This process resulted in a final set of 40, 12, 4, and 33 ToAs at center frequencies of 1369 MHz, 1374 MHz, 1465 MHz, and 3100 MHz, respectively.
\end{itemize}

In the subsequent timing analysis, we use \texttt{TEMPO2} \citep{EdwardsHM2006,HobbsEM2006} to fit these ToAs in order to obtain the rotational parameters and timing residuals.
This process incorporated the Jet Propulsion Laboratory's planetary ephemeris DE440 \citep{ParkFWB2021} and Barycentric Coordinate Time (TCB) to convert the local ToAs to the solar system barycenter.
In addition, to correct for the potential effects of different terminals on the timing results, we introduced additional error-scaling (\texttt{EFAC}) and error-additive (\texttt{EQUAD}) parameters to account for instrumental noise systematics.
The methods for determining these two parameters, along with further technical details can be found in our previous works, e.g., \citet{LiuYGYZ2024,LiuWYT2025,2025MNRAS.537.1720L}.
According to the timing model, the evolution of the pulse phase $\phi(t)$ with time can be expressed as:
\begin{equation}
\label{equ:1} 
\phi(t) = \phi_{\rm 0} + \nu(t - t_{\rm 0}) + \frac{\dot{\nu}}{2} (t - t_{\rm 0})^{2} + \frac{\ddot{\nu}}{6}(t - t_{\rm 0})^{3} \ ,
\end{equation}
where $\phi_{\rm 0}$ is the pulse phase at reference epoch $t_{\rm 0}$. 
$\nu$, $\dot{\nu}$, $\ddot{\nu}$ are the pulsar spin frequency and its derivatives, respectively.
To model the rotational evolution across a glitch, the phase model is augmented with terms describing the instantaneous changes at the glitch epoch $t_{\rm g}$ and the subsequent recovery.
When the post-glitch recovery process is only linear relaxation, the pulse phase change caused by this event can be expressed as:
\begin{equation}
\label{equ:2} 
\phi_{\rm g} = \Delta\phi+ \Delta\nu(t - t_{\rm g}) +   \frac{1} {2} \Delta\dot{\nu} (t - t_{\rm g})^{2}  \ .
\end{equation}
Here, $t_{\rm g}$ represents the epoch of the glitch occurs.
The $\Delta\phi$, $\Delta\nu$ and $\Delta\dot{\nu}$ represent the changes in pulse phase, spin frequency, and the first derivative of spin frequency, respectively.

\begin{table}[ht]
\large
\caption{The timing solutions during the glitch fitting window and glitch parameters obtained for PSR J1718$-$3718.
\label{tab:gl}}  
\begin{flushleft} 
\hspace*{-\dimexpr\columnsep-\arrayrulewidth\relax}
\begin{threeparttable}
    \renewcommand{\arraystretch}{1.2}
    \setlength{\tabcolsep}{10pt}    
\resizebox{0.52\textwidth}{!}{%
\begin{tabular}{lc}
\hline \hline
Parameter                                       & Glitch 1 \\
\hline
Pulse frequency, $\nu$ (Hz)                     & 0.2959835682(4)  \\
Pulse frequency derivative, $\dot{\nu}$ ($10^{-13}$\,s$^{-2}$)  
                                                & $-$1.3999(2) \\
Pulse frequency second derivative, $\ddot{\nu}$ ($10^{-24}$\,s$^{-3}$)  
                                                & $-$4.0(4) \\
Epoch of frequency determination (MJD)          & 54380      \\
Data span (MJD)                                 & 53664--55904  \\
ToA numbers                                     & 42      \\
RMS timing residual ($\rm \mu$s)                & 17009         \\
Glitch epoch (MJD)                              & 54620(240)  \\
$\Delta\nu$ ($10^{-6}$\,Hz)                     & 9.8543(8) \\
$\Delta\nu/\nu$ ($10^{-6}$)               & 33.293(3) \\
$\Delta\dot{\nu}$ ($10^{-15}$\,s$^{-2}$)        & $-$1.00(5) \\
$\Delta\dot{\nu}/\dot{\nu}$ ($10^{-3}$)         & 7.1(3) \\
\hline  \hline 
\end{tabular}}
\end{threeparttable}
\end{flushleft}%
\vspace{-0.2cm}
\end{table}

\section{Timing Results}
\label{sec:result}

In this section, we present a detailed analysis for the timing data of PSR J1718$-$3718 during MJD 51915--58115. By fitting the ToAs to Eq. (\ref{equ:1}) using \texttt{TEMPO2}, we obtained a timing solution that includes parameters $\nu$, $\dot{\nu}$ and $\ddot{\nu}$, and calculated the corresponding timing residuals. The results show that after MJD 54380, the timing residuals exhibit a significant phase discontinuity, a typical characteristic of a large glitch event. After introducing glitch parameters into the timing model, we refitted the ToAs and obtained the timing residuals over the entire time span, as shown in Fig. \ref{fig:residual}.
The figure clearly shows that the timing residuals exhibit significant red noise and have multiple observational gaps, among which there are up to four gaps with time spans exceeding one year.

Table \ref{tab:int} provides the detailed pre- and post-glitch timing solutions, with parameter uncertainties corresponding to the 1$\sigma$ values obtained from \texttt{TEMPO2}. 
The glitch event was first reported by \cite{ManchesterH2011}, with a glitch epoch at MJD 54610(244) and a corresponding $\Delta \nu/\nu \sim 33.25(1) \times 10^{-6}$.
This work uses an extrapolation method to estimate the glitch epoch, which takes the central epoch ($(\rm{t1}+\rm{t2})/2$) between the last observation epoch ($\rm{t1}$) of the pre-glitch and the first observation epoch ($\rm{t2}$) of the post-glitch as the glitch epoch. The uncertainty is defined as half the interval between these two observations ($(\rm{t2}-\rm{t1})/2$). Accordingly, we obtained a glitch epoch of MJD 54620(240).
By further fitting the ToAs to Eq. (\ref{equ:2}) using \texttt{TEMPO2}, we obtained that the glitch size is $\Delta \nu/\nu \sim 33.293(3) \times 10^{-6}$, and the corresponding fractional change in spin-down rate is $\Delta \dot{\nu}/\dot{\nu} \sim 7.1(3)\times10^{-3}$, which are generally consistent with the results reported by \cite{ManchesterH2011}.
More glitch parameters are listed in Table \ref{tab:gl}, where the uncertainty of the parameters in rows 11 and 13 were obtained by error propagation equation, while the uncertainties for the remaining parameters are taken as the 1$\sigma$ values obtained from \texttt{TEMPO2}.

\begin{figure}
\centering
\includegraphics[width=0.45\textwidth]{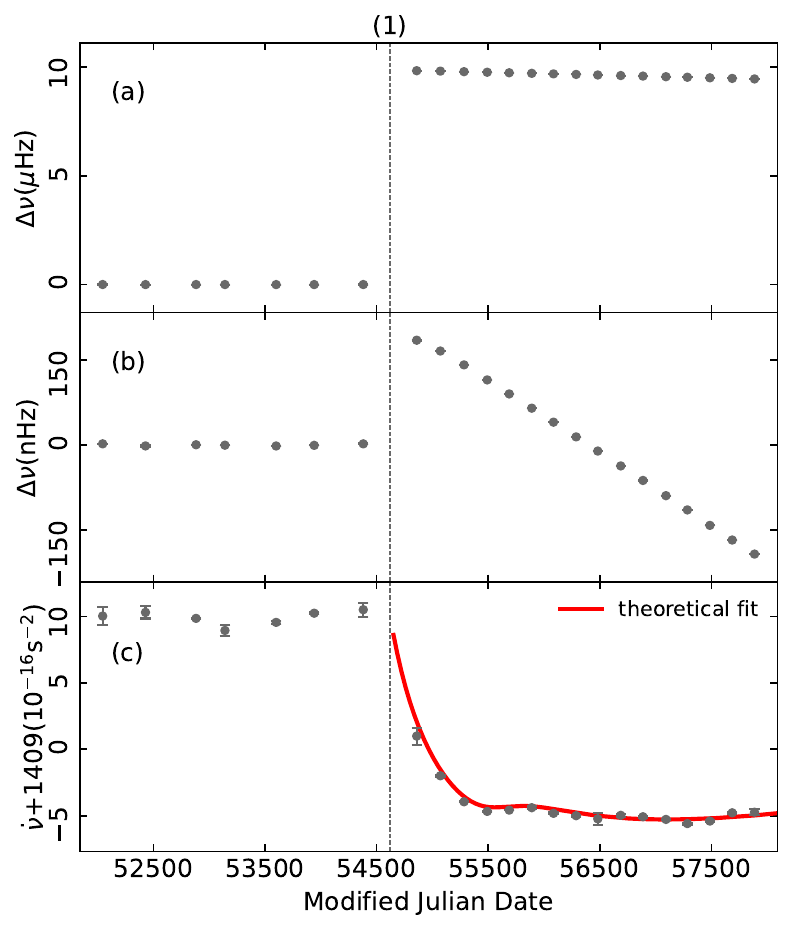}
\caption{Glitch in PSR J1718$-$3718: Panel (a) shows the relative change of spin frequency ($\Delta\nu$), that is, the change in pulsar rotation frequency relative to the frequency model before the glitch. Panel (b) is an enlarged view of the post-glitch $\Delta\nu$ evolution, centered on its mean value to highlight the recovery details.
Panel (c) shows the evolution of the spin-down rate ($\dot{\nu}$) over time before and after the glitch. The vertical grey-dashed line and the numbers in between parentheses at the top represent the glitch epoch and the reference number of glitch occurrences, respectively. 
In Panel (c), the red curves represent the fit using the superfluid glitch model (see Section \ref{sec:fits}), with the corresponding fit parameters provided in Table \ref{tab:posterior}.
}
\label{fig:J1718-3718}
\end{figure}

To investigate the pulsar's spin behavior over the entire data span, we divided the ToAs into a series of overlapping time windows and performed separate fits, thereby deriving the $\nu$ and $\dot{\nu}$ at different epochs. Each window spans approximately 800 d, with an overlap of about 600 d between adjacent windows. Given the low observing cadence and the significant observational gaps before the glitch, we have appropriately adjusted the time spans of some windows. Fig. \ref{fig:J1718-3718} shows the evolution of $\nu$ and $\dot{\nu}$ over time for PSR J1718$-$3718. The figure clearly indicates that there is a significant jump in the spin of the pulsar between MJD 54380 and 54860.
Among them, panels (a) and (c) show changes in spin frequency of $\Delta \nu \sim 10 \, \rm{\mu Hz}$ and the change in spin-down rate of $\Delta \dot{\nu} \sim -10^{-15} \, \rm{s^{-2}}$, respectively, which are consistent with the fitted parameters listed in Table \ref{tab:gl}.
Unlike typical glitches where $\dot{\nu}$ recovers exponentially toward its pre-glitch value, Panel (c) shows that $\left| \dot{\nu} \right|$ increases exponentially for approximately 900 days post-glitch before stabilizing. 
This unusual behaviour is the primary focus of our modeling.

\begin{table*}[htbp]
  \centering
  \caption{Proposed mechanisms for explaining the post-glitch rotational evolution of PSR~J1718$-$3718.
  }
  \label{tab:compoent}
\begin{tabularx}{\textwidth}{ 
    >{\RaggedRight}p{3.5cm}
    >{\RaggedRight}p{4.5cm}
    >{\RaggedRight}p{3.5cm}
    >{\RaggedRight\arraybackslash}X
}
\toprule
 & \textbf{Mechanism} & \textbf{Observational Signatures} & \textbf{Relevant Physics of PSR J1718$-$3718 }  \\
\midrule
\textbf{Vortex outward creep} &
Quantum tunneling between adjacent pinning sites driven by Magnus forces in the outward direction &
Decreased external torque and $\dot{\nu}$; \newline
Linear response: non-linear relaxation; \newline
Nonlinear response: approximate linear relaxation &
Internal temperature; \newline
Nuclear pinning force; \newline
Mol involved in outward creep, $\log_{10}I_{\rm{nl}}/I\sim-2.616$. \\
\midrule
\textbf{Vortex inward motion} &
External agent acts directly on the superfluid  (e.g.,
starquake-induced motion) &
Increased external torque and $\dot{\nu}$ &
Mol involved in inward motion, $\log_{10}I_{\text{i}}/I\sim-2.156$;\newline
Stellar Mol change, $|\Delta I/I|\gtrsim1.452\times10^5$; \newline
Energy dissipation due to crust breaking, $\gtrsim2.753\times10^{42}$ erg; \newline
Estimated number of crustal plates, $\sim142$; \newline
Typical Plate size, $\sim0.03$ km \\
\midrule
\textbf{External torque variation} & 
Magnetic dipole radiation;
\newline
Change in inclination angle; \newline Variation in stellar wind
strength \newline &
Braking index deviation; \newline
Pulse profile change &
Surface magnetic field configuration; \newline 
Plasma gap location; \newline
Wind particle density. \\
\bottomrule
\end{tabularx}
\end{table*}

\section{Model Fits}
\label{sec:fits}

\subsection{The Vortex Creep Model with Inward Motion}

The vortex creep model explains glitches as sudden angular momentum transfers from a superfluid neutron star interior to the solid crust~\citep{1984ApJ...276..325A,1989ApJ...346..823A,1993ApJ...409..345A}. 
The moments of inertia (Mol) of the superfluid components involved in glitches are key to understanding glitch sizes and activities~\citep[e.g.,][]{2016ApJS..223...16L,2025ApJ...984..200T,2025arXiv251020791T}.
In the inner crust, neutron superfluid vortices can be pinned to nuclei. Under the spin-down of the crust, a velocity lag $\omega$ develops between the superfluid and the crust. Vortices can slowly $creep$ outward through thermal activation and quantum tunneling, transferring angular momentum to the crust and sustaining the steady-state spin-down ($\omega_{\infty}$).

A glitch is triggered when a sudden avalanche of unpinned vortices moves outward, rapidly spinning up the crust. The post-glitch relaxation is governed by how different superfluid regions respond to the resulting perturbation in the lag, $\delta\omega=\omega-\omega_{\infty}$. The standard model, considering only outward vortex motion, predicts a post-glitch decrease in the spin-down rate ($|\dot{\nu}|$), which contradicts the observed increase in PSR J1718--3718. To explain this, we incorporate vortex inward motion. While outward creep is driven by the Magnus force, inward motion requires an external agent, such as a crustquake~\citep{1991ApJ...366..261R}.

A crustquake can suddenly shift crustal plates, creating a deficit of vortices in certain regions. To re-establish equilibrium, vortices in adjacent regions move inward. This inward motion transfers angular momentum to the superfluid, effectively applying an additional brake on the crust and increasing $|\dot{\nu}|$. This extended framework, incorporating both outward and inward vortex motion, provides a plausible mechanism for the unique post-glitch behavior of PSR J1718--3718, as demonstrated in analyses of other pulsars like PSR J1119--6127~\citep{2015MNRAS.449..933A}.
The three physical components contributing to the post-glitch rotational evolution—vortex outward creep, vortex inward motion, and external torque variations—are summarized in Table \ref{tab:compoent}, which details their underlying mechanisms, observational signatures, and the physical parameters constrained for PSR J1718$-$3718.

\subsection{Fitting Method and MCMC Setup}

\begin{figure*}
\centering
\includegraphics[width=1.0\textwidth]{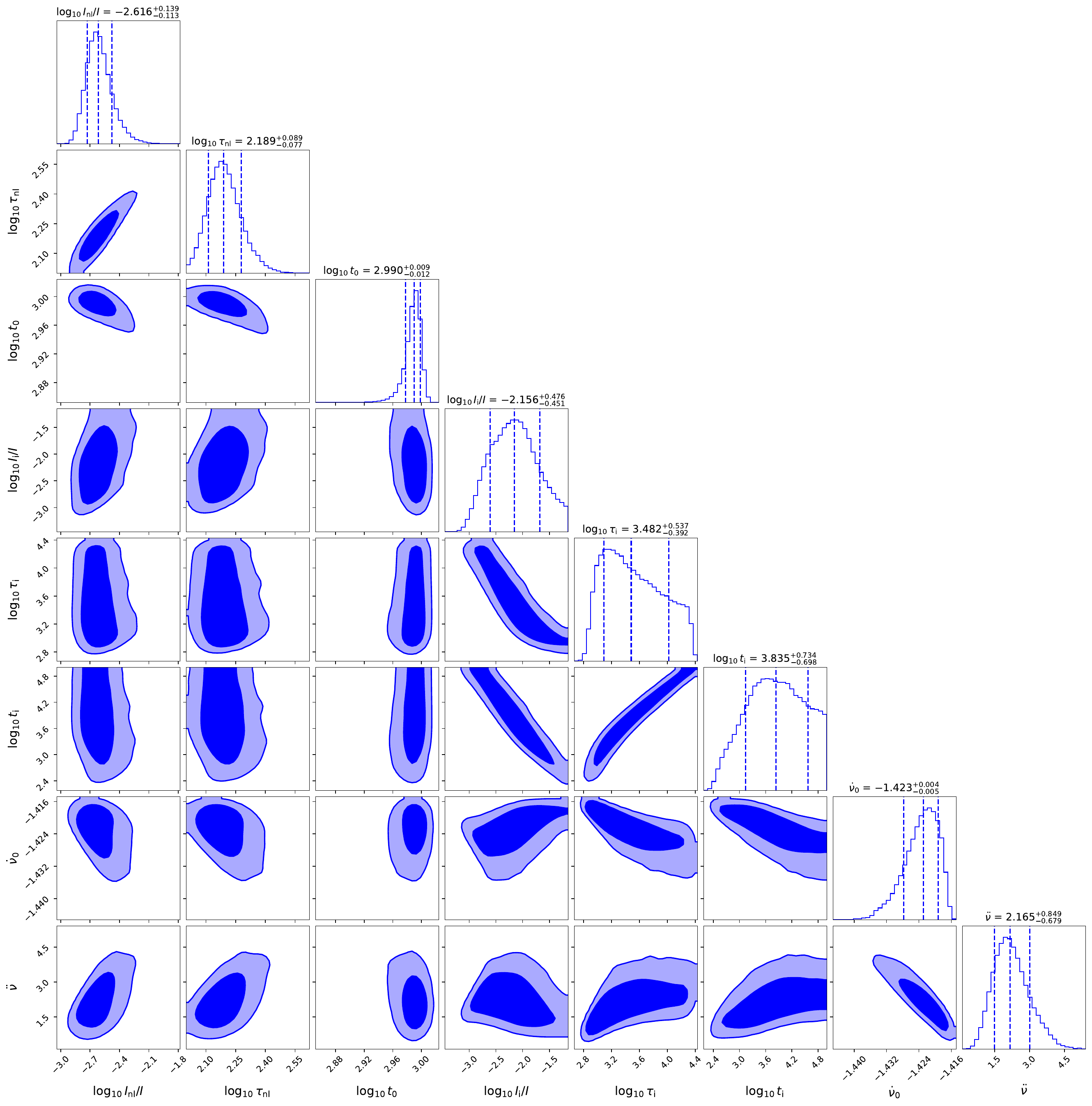}
\caption{Posterior distributions of the fitting parameters of the vortex creep model for PSR J1718--3718 glitch. 
The diagonal panels show the marginalized posterior distributions for each parameter, while the off-diagonal panels show the 2D correlations between parameters.
The crustal fractional MoIs associated with vortex outward and inward creep are $I_{\mathrm{nl}}/I$ and $I_{\mathrm{i}}/I$, with the corresponding creep relaxation timescales being $\tau_{\mathrm{nl}}$ and $\tau_{\mathrm{i}}$, respectively. 
While $t_0$ represents the waiting time for outward vortex creep, the parameter $t_{\mathrm{i}}$ characterizes the initial magnitude of the lag perturbation that drives the inward vortex motion, related to the crustquake's intensity.
Both the fractional Mol and the time parameters are presented on a logarithmic scale in the figure. $\dot{\nu}_0$ and $\ddot{\nu}$ are the first- and second-order derivatives of spin frequency driven by an external mechanism (e.g., the pulsar wind model), with units of $10^{-13}$ Hz/s and $10^{-24} \rm{Hz/s}^2$, respectively.
}
\label{fig:Posterior}
\end{figure*}

We model the post-glitch recovery using vortex outward/inward motion plus external torque evolution ($\dot{\nu}_{\mathrm{ex}}=\dot{\nu}_0+\ddot{\nu}t$).  
After performing preliminary fitting via the least-squares method, we derived an expression of post-glicth rotational evolution for further refinement:
\begin{align}
    \frac{\Delta\dot{\nu}_{\mathrm{c}}(t)}{\dot{\nu}_\mathrm{c}}&=-\frac{I_{\rm{nl}}}I\left\{1-\frac{1}{1+(e^{t_{\mathrm{0}}/\tau_{\mathrm{nl}}}-1)e^{-t/\tau_{\mathrm{nl}}}}\right\}, \notag \\
    &\quad +\frac{I_\mathrm{i}}{I}\left\{1-\frac{1+(\tau_\mathrm{i}/t_\mathrm{i})\ln\left[1+(\mathrm{e}^{-t_\mathrm{i}/\tau_\mathrm{i}}-1)\mathrm{e}^{-\frac{t}{\tau_\mathrm{i}}}\right]}{1-\mathrm{e}^{-\frac{t}{\tau_\mathrm{i}}}}\right\} \notag \\
    &\quad + \dot{\nu}_0+\ddot{\nu}t.
    \label{eq:fitting}
\end{align}
Here, $\tau_{\rm{nl}}$ and $I_{\rm{nl}}/I$ are the relaxation time and MoI of a single non-linear response region.  $\tau_{\rm{i}}$ and $I_{\rm{i}}/I$ are the relaxation time and fractional MoI of several neighboring single vortex inward motion region. $t_0$ and $t_{\rm{i}}$ relate to $\delta\omega=2\pi\delta\nu$ by $t_0=\delta\nu_0/|\dot\nu|_{\infty}$ and $t_0=\delta\nu_{\rm{i}}/|\dot\nu|_{\infty}$ with the lag deviations $\delta\nu_0$ and $\delta\nu_{\rm{i}}$ of vortex outward and inward motion regions, respectively.

Since no linear response signatures of outward vortex creep were observed and the least-squares fitting demonstrated extreme insensitivity of the results to the parameters of the linear response term, we omitted the linear response term of outward vortex creep in Eq. (\ref{eq:fitting}). Given that there are currently only 17 data points during the glitch recovery process, least-squares fitting cannot produce accurate error estimates; we also aim to analyze the parameter degeneracy within and between various components during the post-glitch recovery.  We run a Markov chain Monte Carlo fitting for the PSR J1718$-$3718 glitch event by using \texttt{emcee} \citep{2013PASP..125..306F}. The uniform distributions are adopted as the priors for all fitting parameters, with the parameter ranges determined by the least-squares fitting. Previous studies have indicated that the total MoI involved in the glitch event does not exceed 7\% of the crustal MoI \citep{2012PhRvL.109x1103A,2013PhRvL.110a1101C}, we apply the parameter constraint $(I_{\mathrm{nl}}+I_{\mathrm{i}})/I<0.07$ in our fitting. The fractional MoI and time parameters (e.g., $\tau_{\mathrm{nl}}$ and $t_{\mathrm{0}}$) can span two to three orders of magnitude, e.g., the fractional MoI ranges from $10^{-4}$ to $0.07$, therefore logarithmic parameterization is applied to these parameters before performing MCMC sampling. The Gaussian likelihood is adopted for our fitting.

\subsection{Fitting Results and Parameter Estimates}

Fig. \ref{fig:Posterior} presents the posterior distribution fitted to the recovery process of the glitch event in PSR J1718--3718. The parameter estimations and their 1$\sigma$ errors are listed in Table \ref{tab:posterior}. The 1$\sigma$ confidence intervals of $\log_{10}I_{\mathrm{nl}}/I$ and $\log_{10}I_{\mathrm{i}}/I$ fall within reasonable ranges, specifically below $\log_{10}0.07\approx-1.15$. The unit of time parameters is day. The fitted curve calculated by the median values is presented in Fig. \ref{fig:J1718-3718} to compare with observations. We can see that the fitted curve agrees well with the observations. Both the vortex inward creep and outward creep induced by the star quake collectively contribute to angular momentum transfer during glitch triggering. Additionally, the reduction in stellar MoI caused by the starquake also contributes to angular momentum transfer, leading to a spin-up $\Delta\nu_{\mathrm{c}}/\nu_{\mathrm{c}}=-\Delta I/I$, as required by angular momentum conservation. Therefore, the angular momentum conservation during a glitch event is given by:
\begin{align}
    I\Delta\nu_c = (I_{\mathrm{nl}}+I_{\mathrm{B}})\Delta\nu_{\mathrm{s}}-(I_{\mathrm{i}}/2+I_{\mathrm{B}})\Delta\nu_{\mathrm{s}}^{\prime}-\Delta I\nu_{\mathrm{c}}, \label{eq:amc}
\end{align}
where $\Delta\nu_{\mathrm{s}}\approx|\dot{\nu}|_{\infty}t_{\mathrm{0}}$ and $\Delta\nu_{\mathrm{s}}^{\prime}\approx|\dot{\nu}|_{\infty}t_{\mathrm{i}}$, $I_{\rm{B}}$ is the MoI associated with the region through which vortices repin from one non-linear response region to another after unpinning. We have assumed that inward moving vortices traversed the same radial distance as those moved outward, $I_{\mathrm{B}}$ is the same for the inward and outward vortex creeps. We obtain $|\Delta I/I|>1.452\times10^{-5}$ for $I_{\mathrm{B}}>0$, the energy dissipation due to crust breaking is estimated as $\Delta E_{\mathrm{elastic}}=I\Omega_{\mathrm{c}}^2|\Delta I/I|/2\gtrsim2.753\times10^{41}$ erg if we adopt $I=10^{46}~\mathrm{g~cm}^2$.

Next, we analyze the degeneracies between different parameters. We find that there is no strong parameter degeneracy between the inward and outward vortex creep components, which suggests that the two mechanisms are relatively independent in this glitch event. However, each component shows strong internal correlations. For the outward vortex creep, angular momentum conservation, i.e. Eq. (\ref{eq:amc}), clearly shows that $I_{\mathrm{nl}}$ and $t_{\mathrm{0}}$ is anti-correlated; $\tau_{\mathrm{nl}}$ and $t_{\mathrm{0}}$ are also anti-correlated because they are used to fit the timescale of the outward vortex creep, namely $t_{\mathrm{0}}+3\tau_{\mathrm{nl}}$; these two parameter dependencies directly lead to $I_\mathrm{nl}$ and $\tau_\mathrm{nl}$ being positively correlated. These correlations have been indicated in previous studies \citep{Grover2025_arXiv2506.02100}. In Eq. (\ref{eq:fitting}), $t_{\mathrm{i}}$ does not function as a waiting time like $t_0$, hence the nonlinear response of inward vortex creep exhibits quasi-exponential relaxation and $I_{\mathrm{i}}$ and $\tau_{\mathrm{i}}$ are anti-correlated; similarly, angular momentum conservation leads to $I_{\mathrm{i}}$ and $t_{\mathrm{i}}$ being anti-correlated; these two parameter correlations directly lead to $\tau_{\mathrm{i}}$ and $t_{\mathrm{i}}$ being positively correlated. The component $\dot{\nu}_{\mathrm{ex}}$ associated with external torques needs to be adjusted under changing vortex creep components to fit post-glitch recovery, thus exhibiting correlations with both vortex creep components. The correlations between different components indicate that both the triggering of the glitch and the subsequent post-glitch recovery result from the combined effects of several mechanisms.
Despite these internal correlations, the inward and outward creep components are independent, indicating they originate from distinct physical regions within the star.

Our model suggests that a crustquake was the likely trigger for the glitch. Such an event, involving the motion of crustal plates, can also reconfigure the neutron star's magnetosphere by altering the surface magnetic field~\citep{2015MNRAS.449..933A}, the location of plasma gaps, or the particle wind density. These magnetospheric changes would manifest as a long-term evolution of the external braking torque.
To quantitatively assess this effect, we incorporate a time-varying external torque within the framework of the pulsar wind model. For simplicity, we assume the magnetic inclination angle $\alpha$ remains constant and attribute the torque variation solely to a change in the particle wind density parameter, $\kappa$. The rotational evolution is then described by the pulsar wind model:
\begin{equation}\label{pulsar_wind_model}
    -I\Omega\dot{\Omega}=\frac{2\mu^2\Omega^4}{3c^3}\eta,
\end{equation}
where $\mu = BR^3/2$ is the magnetic dipole moment with the polar magnetic field $B$ and the radius of pulsar $R$, $c$ is the speed of light. $\eta$ is the particle acceleration model, we adopt the VG(CR) model,
\begin{equation}\label{VGCR}
    \eta = \sin^2{\alpha}+496\kappa B_{12}^{-8/7}\Omega^{-15/7},
\end{equation}
where $B_{12}$ is the magnetic field strength in units of $10^{12}$ G, $\kappa$ is the ratio of particle density to the primary particle density. The weighting factor $\cos^2\alpha$ in the particle wind is omitted due to the results of the magnetospheric simulations \citep{2017ApJ...837..117T}. For PSR J1718--3718, $B_{12}=74.7$; we adopt $I = 10^{46}$ g cm$^2$ and $R=12$ km in our fitting. Using the values of $\nu$ and $\dot{\nu}$ just before the glitch, we obtain $\kappa\approx$13-14 and simply take $\kappa=$13.5. This gives us $\alpha\approx40.73^\circ$. We assume that $\alpha$ is unchanged ($\Delta\alpha$ is small in the 2007 glitch of PSR J1119-6127) but $\kappa$ changes with time after the glitch due to the change in structure of the pulsar, $\kappa(t)=\kappa_0+\dot{\kappa}t$. Combining the fitting results, we obtain $\kappa_0\approx13.786$ and $\dot{\kappa}\approx-2.165\times10^{-10}~\rm{s}^{-1}$. The change in $\kappa$ from 13.5 to 13.786 indicates a possible redistribution of crustal matter.

The number of unpinned vortices $\delta N_{\mathrm{out}}$ which moved outward during a glitch is related to the decrease $\Delta\nu_{\mathrm{s}}$. From our fitting, $\delta N_{\mathrm{out}}=2\pi R^2\frac{2\pi\Delta\nu_{\mathrm{s}}}{\kappa_{\mathrm{v}}}\approx3.394\times10^{11}$, where $\kappa_{\mathrm{v}}=\pi\hbar/m_{N}$ is the quantum of vorticity. Similarly, the number of pinned vortices $\delta N_{\mathrm{in}}$ which moved inward with all broken plates is related to the increase $\Delta\nu_{\mathrm{s}}^{\prime}$, $\delta N_{\mathrm{in}}=2\pi R^2\frac{2\pi\Delta\nu_{\mathrm{s}}^{\prime}}{\kappa_{\mathrm{v}}}\approx2.375\times10^{12}$. The broken plate size $D$ in a crustquake can be estimated by using simple geometrical arguments \citep{2015MNRAS.449..933A}
\begin{align}
    I_{\mathrm{i}}/I\simeq\frac{4\pi\rho_{s}R^4D\sin{\alpha}\cos^2{\alpha}}{(2/5)MR^2}\approx\sin{\alpha}\cos^2{\alpha}\frac{15D}{2R},
\end{align}
Substituting $\alpha\approx40.73^\circ$ and $R=12$ km, we obtain $D\approx0.03$ km. Then the number of vortices pinned to one broken plate is estimated as $\delta N_{\mathrm{plate}}\simeq 4\pi\nu_{\mathrm{c}}D^2/\kappa_{\mathrm{v}}\approx1.672\times10^{10}$. Therefore the number of plates involved in the glitch is $\delta N_{\mathrm{in}}/\delta N_{\mathrm{plate}}\sim142$, which is one order of magnitude smaller than that infered from the largest Crab glitch, that is, $\delta N_{\mathrm{in}}/\delta N_{\mathrm{plate}}\sim10^3$ \citep{2019MNRAS.488.2275G}. This implies that the breaking strain of PSR J1718--3718's crust is larger than that of Crab. Recent molecular dynamics simulations of crust breaking also suggested a large breaking strain $\sim0.1$ \citep{2009PhRvL.102s1102H}.

\begin{table*} 
\centering
\renewcommand{\arraystretch}{1.35}
\caption{The posterior parameters estimated by fitting the vortex creep model to the observed post-glitch recovery of PSR J1718$-$3718. The fractional MoIs and creep timescales are sampled after logarithmic parameterization. The unit of creep timescale is day. $\dot{\nu}_0$ and $\ddot{\nu}$ are in units of $10^{-13}~\mathrm{Hz/s}$ and $10^{-24}~\mathrm{Hz/s^2}$.
\label{tab:posterior}}  
\vspace{-0.3cm}
\setlength{\tabcolsep}{4.5pt}  
\begin{tabular}{lcccccccc}
\hline\hline 
PSR &$\log_{10}I_{\mathrm{nl}}/I$ &$\log_{10}\tau_{\mathrm{nl}}$ &$\log_{10}t_{\mathrm{0}}$ &$\log_{10}I_{\mathrm{i}}/I$ &$\log_{10}\tau_{\mathrm{i}}$ &$\log_{10}t_{\mathrm{i}}$ &$\dot{\nu}_{0}$ &$\ddot{\nu}$  \\ 
 & & & & 
& & & ($10^{-13}~\mathrm{Hz/s}$) &($10^{-24}~\mathrm{Hz/s^2}$) \\ 
\hline
J1718$-$3718  & $-2.616^{+0.139}_{-0.113}$ & $2.189^{+0.089}_{-0.077}$ & $2.990^{+0.009}_{-0.012}$ & $-2.156^{+0.476}_{-0.451}$ & $3.482^{+0.537}_{-0.392}$ & $3.835^{+0.734}_{-0.698}$ & $-1.423^{+0.004}_{-0.005}$ & $2.165^{+0.849}_{-0.679}$  \\ 
\hline\hline 
\end{tabular}
\end{table*}

Our fitting results suggest that the pulsar structure changes before and after the glitch, indicating that this glitch is likely triggered by starquake. The specific post-glitch recovery mechanism still requires short-term observations following the glitch to be distinguished.

\section{Conclusions} 
\label{sec:Conclusions}

In this work, we investigated the spin evolution behavior of PSR J1718$-$3718 over a 17 yr span using timing observation data obtained with Murriyang, the Parkes radio telescope between 2000 and 2017.
We updated the relevant parameters of the large glitch that occurred at MJD 54620(240) and confirmed the subsequent recovery process: the $\left| \dot{\nu} \right|$ increased exponentially over a relaxation timescale of approximately 900 d, after which it gradually approached a stable state.

In general, following a large glitch, the $\dot{\nu}$ of a pulsar tends to evolve toward its pre-glitch state.
The typical glitch recovery phase begins with an exponential relaxation process, usually occurring on a timescale of 10 to 300 d, and subsequently transitions into a linear relaxation phase that can last for several years \citep{YuanWML2010}.
However, there are also some pulsars that enter the linear relaxation process immediately after the glitch \citep{BasuSAK2022}.
Among the reported large glitch events, PSR J1718$-$3718 is the only pulsar whose $\left| \dot{\nu} \right|$ exhibits an exponential increase within a few hundred days after the glitch.
The standard vortex creep model, which assumes only outward radial vortex motion, predicts a post-glitch decrease in the spin-down rate (i.e., a recovery toward the pre-glitch value). This is incompatible with the observed increase in $\left| \dot{\nu} \right|$ for PSR J1718$-$3718. 

To explain this, we employ an extended model that incorporates vortex inward motion, which can lead to an increased braking torque.
The model parameters and associated uncertainties are quantified using a Markov Chain Monte Carlo approach.  Within the vortex creep model, the exponentially increased $\left| \dot{\nu} \right|$ shows a signal of vortex inward motion. Our model simultaneously accounts for contributions to the rotational evolution from both vortex outward and inward motion, while explicitly incorporating the effects of external torque. The results reproduce the observed post-glitch recovery and indicates that the vortex inward creep and a long-term change in external torque dominate the observed increase in spin-down rate, suggesting structural changes triggered by a crustquake that initiated both vortex motion and a change in the moment of inertia. We estimate that the glitch involved approximately $3.4 \times 10^{11}$ outward-moving vortices, and a total of $2.4 \times 10^{12}$ inward-moving vortices distributed across  $\sim142$ crustal plates with a typical size of $\sim0.03$ km. 

Our analysis of PSR J1718$-$3718 demonstrates that giant glitches can be powerful probes of neutron star interiors, revealing not just superfluid dynamics but also crustal rigidity and magnetospheric coupling. Future high-cadence observations of similar events will be crucial for further testing and refining this unified framework.

\section*{Acknowledgments}
The work is supported by the National Natural Science Foundation of China (grant Nos. 12041304, 12273028, 12494572) and the National SKA Program of China (No.~2020SKA0120300).
Murriyang, CSIRO’s Parkes radio telescope, is part of the Australia Telescope National Facility, which is funded by the Commonwealth of Australia for operation as a National Facility managed by CSIRO. This paper includes archived data obtained through the CSIRO Data Access Portal.

\bibliographystyle{aasjournal}

\end{document}